\documentclass[referee]{raa}            

\usepackage{graphicx,times}             
\usepackage{natbib}
\usepackage{amssymb,amsmath}
\bibpunct{(}{)}{;}{a}{}{,}

\usepackage[a4paper=true,dvipdfm=true,pagebackref=true]{hyperref}
\hypersetup{colorlinks = true, linkcolor = green, anchorcolor =
red, citecolor = blue, filecolor = red, pagecolor = red, urlcolor
= red}

\begin{document}

   \title{On the metallicity gradient in the galactic disk
}

   \volnopage{Vol.0 (20xx) No.0, 000--000}      
   \setcounter{page}{1}          

   \author{A. V. Loktin
      \inst{1}
   \and M. E. Popova
      \inst{1}
   }

   \institute{Astronomical Observatory, Ural Federal University,
             Ekaterinburg, Russia; {\it maria.popova@urfu.ru}\\
\vs\no
   {\small Received~~20xx month day; accepted~~20xx~~month day}}

\abstract{ The problem of the chemical composition gradient in the
galactic disk is studied based on a sample of metallicity
estimates of open star clusters, using Gaia DR2-improved distance
estimates. A clearly non-monotonic variation was observed in the
average metallicity of clusters with increasing galactocentric
distance. One can clearly see the metallicity jump of $0.22$ in
$[Fe/H]$ at a Galactocentric distance of about $9.5$~kpc, which
appears to be linked to the outer boundary of the thinnest and
youngest component of the galactic disk. The absence of a
significant metallicity gradient in the internal ($R<9$~kpc) and
external ($R>10$~kpc) regions of the disk demonstrates the absence
of noticeable metal enrichment at times of the order of the ages
corresponding to those of the disk regions under consideration.
Observational data show that the disk experiences noticeable metal
enrichment only during the starburst epochs. No significant
dependence was found between the average metallicity and the age
of the clusters. }

   \authorrunning{A. V. Loktin \& M. E. Popova }            
   \titlerunning{On the metallicity gradient in the galactic disk }  

   \maketitle

%
%
\section{Introduction}           
\label{sect:intro}

High-precision photometric and astrometric data from the Gaia DR2
catalog \citep{GaiaDR2_2018} have heavily influenced research in a
number of galactic astronomy fields. One such study examines the
spatial gradient of the chemical composition in the Galactic disk.
Another rather interesting subject is the time gradient of the
chemical composition as it relates to the rate of heavy element
enrichment of the disk. These gradients have been studied
previously \citep{Magrini2017, Marsakov2016, Huyan2015,
Kubryk2015, Mikolaitis2014, Marsakov2014, Gozha2013}. References
to earlier studies on the metallic gradient of the galactic disk
can be found in \cite{Gozha2012A}. The \cite{Lepine2011} gives a
detailed analysis of influence of the spiral structure of the
galactic disk on its chemical structure. The Gaia DR2 data has
made it possible to more precisely define the distances to many
stellar objects, giving impetus to our research on this matter. We
decided to study the gradients of the chemical composition based
on open star cluster (~OCls) data, as these objects are observed
at large distances from the Sun, and therefore at greater range of
galactocentric distances. Their ages cover the entire range of
galactic disk ages, which distinguishes these objects as compared
to, for example, classical Cepheids, whose ages are, to a certain
extent, unreliable, as the age range of these massive, obviously
young stars is quite small.


\section{Sample of OCl data}
\label{sect:Samp}

\begin{figure}
  \begin{minipage}[t]{0.495\linewidth}
  \centering
   \includegraphics[width=50mm]{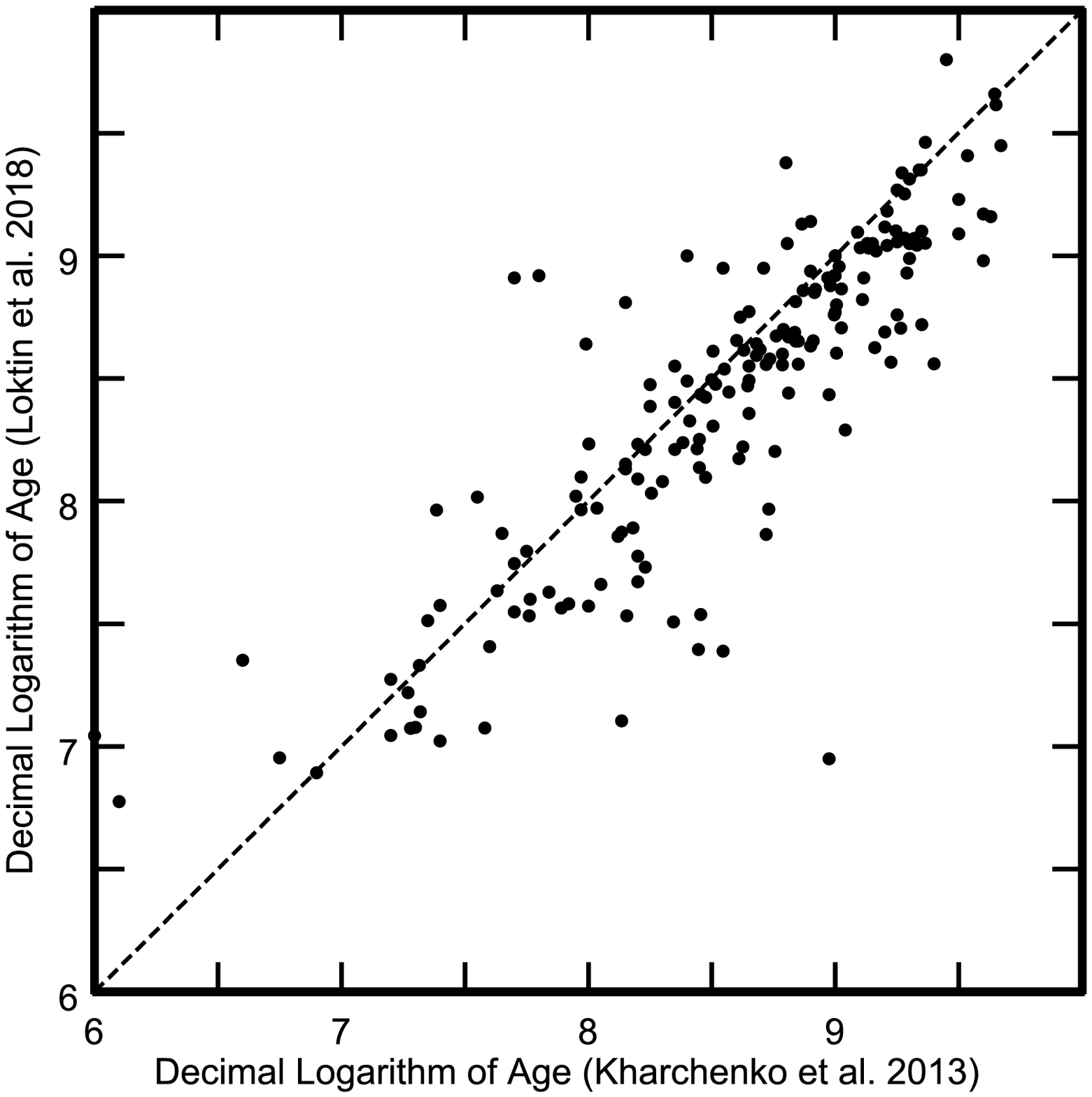}
  \end{minipage}%
  \begin{minipage}[t]{0.495\textwidth}
  \centering
   \includegraphics[width=50mm]{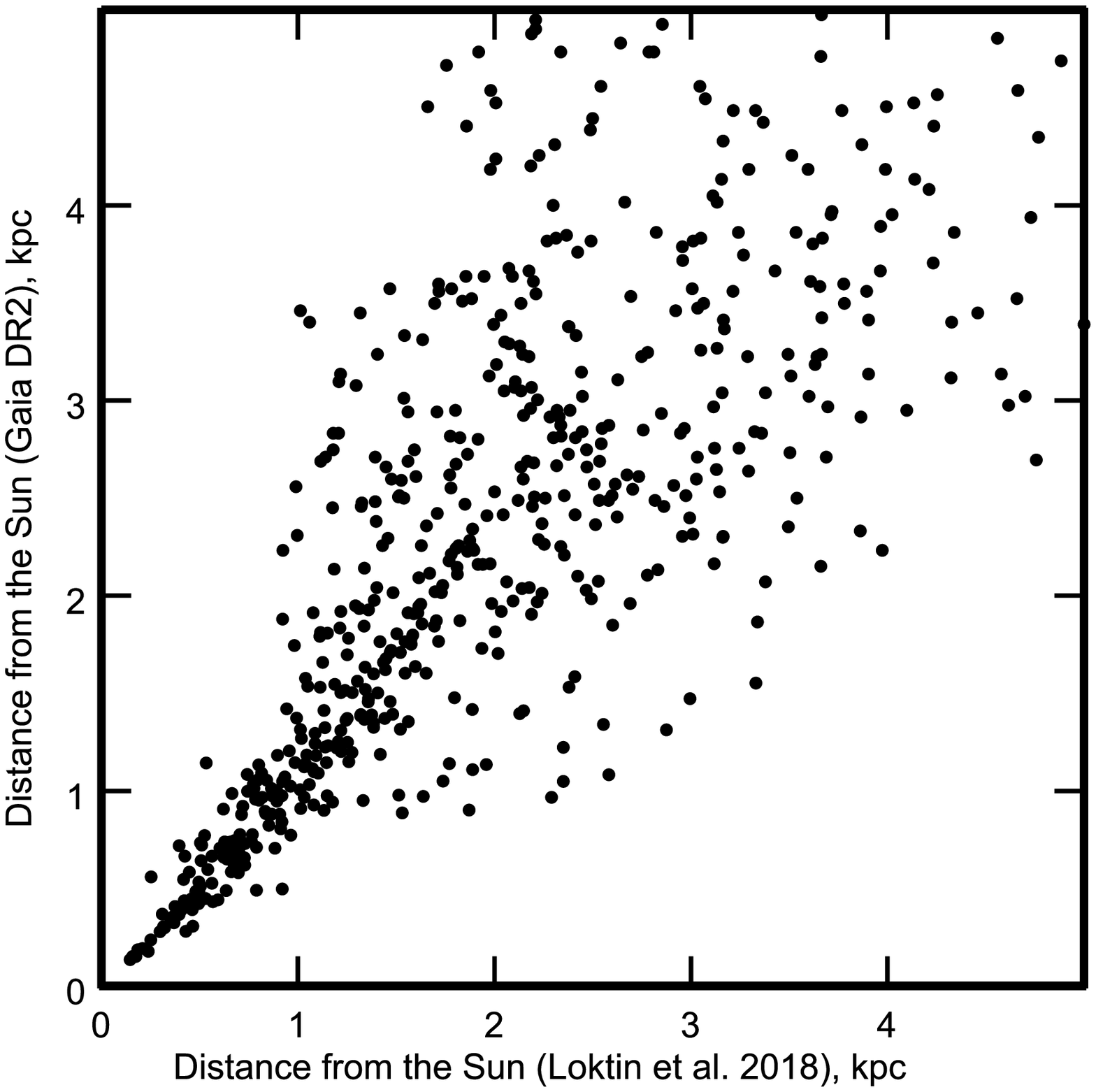}
  \end{minipage}
  \caption{Comparison of OCl ages from the "Homogeneous Catalog of
  Open Cluster Parameters" and the catalog of \cite{Kharchenko2013} (left).
  Comparison of heliocentric distances of OCls from
   the "Homogeneous Catalog of Open Cluster Parameters" and Gaia DR2 data (right).}
  \label{Fig1}
\end{figure}

The "Homogeneous Catalog of Open Cluster Parameters"
\citep{Loktin2017} provided the data on the open clusters. Cluster
metallicities are from the catalogs of \cite{Gozha2012B},
\cite{Kharchenko2013}, \cite{Dias2002}. For those clusters with a
known metallicity not found in the "Homogeneous Catalog of Open
Cluster Parameters", the ages of the OCls were taken from the
catalog of \cite{Kharchenko2013}. Figure~\ref{Fig1} (left) shows a
comparison of OCl ages taken from the "Homogeneous Catalog of Open
Cluster Parameters" and of those from the catalog of
\cite{Kharchenko2013}. The points are positioned close to the
$45^\circ$ line, an unsurprising observation due to the
similarities of the age determination methods and theoretical
isochrones sets used, and therefore supports the inclusion of the
ages from the catalog of Kharchenko et al. in our sample. Not all
clusters with metallicity estimations have distance estimates in
the "Homogeneous Catalog of Open Cluster Parameters". To ensure
greater uniformity in the initial data, the distances to the OCls
in our sample were calculated using parallaxes from the catalog of
\cite{CantatGaudin2018}, which is based on Gaia DR2 data.
Figure~\ref{Fig1} (right) shows a comparison of the OCl distances
from the Sun, determined using trigonometric parallaxes, to the
distance estimates obtained primarily through diagram fitting and
taken from the "Homogeneous Catalog of Open Cluster Parameters".
The figure shows that the distance estimates from both are close
for most clusters located within $1.5$~kpc from the Sun, while the
same cannot be said for more distant clusters. At the same time,
however, errors in the estimates from both the trigonometric
parallaxes and photometry increases at farther distances, as
distant clusters require deep photometry. Errors in determining
color excesses and interstellar absorption also increase as
distance from the Sun increases. The final sample includes
clusters with a deviation from the Galactic plane that did not
exceed $\pm 1.5$~kpc. In total, the sample included $322$ OCls and
contains clusters with decimal logarithms of age ($\log T$)
ranging from $6.6$ to $9.8$, which covers the entire age interval
for galactic disk objects. The galactocentric distances of the
sample cover the interval $5 - 17$~kpc, and the metallicities
$[Fe/H]$ of these OCls fall within the range of $-1.54 - +0.46$.

\section{Metallicity gradients}
\label{sect:grad}

\begin{figure}
  \begin{minipage}[t]{0.33\linewidth}
  \centering
  \includegraphics[width=45mm]{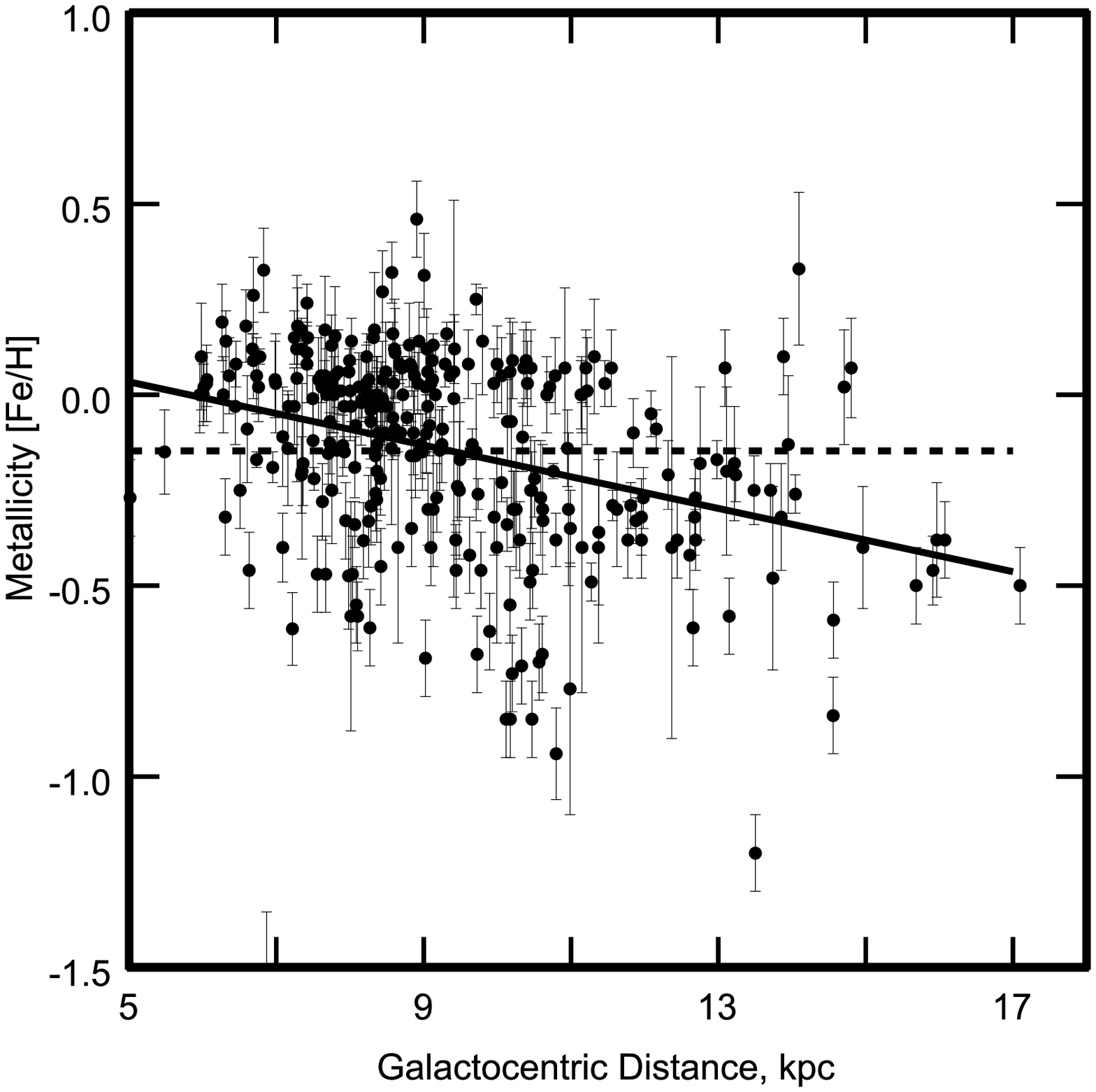}
  \end{minipage}%
  \begin{minipage}[t]{0.33\textwidth}
  \centering
  \includegraphics[width=45mm]{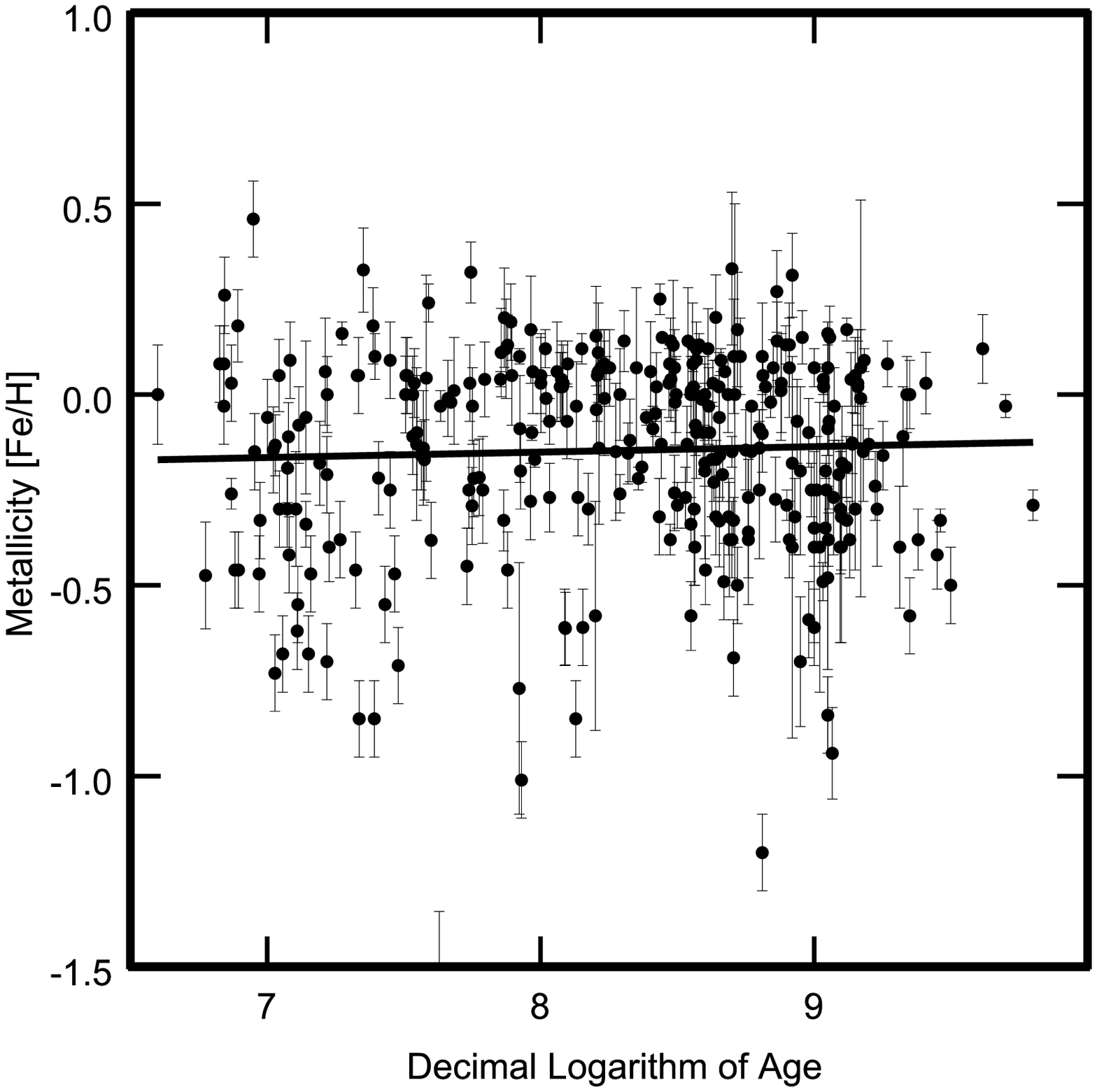}
  \end{minipage}%
  \begin{minipage}[t]{0.33\linewidth}
  \centering
  \includegraphics[width=45mm]{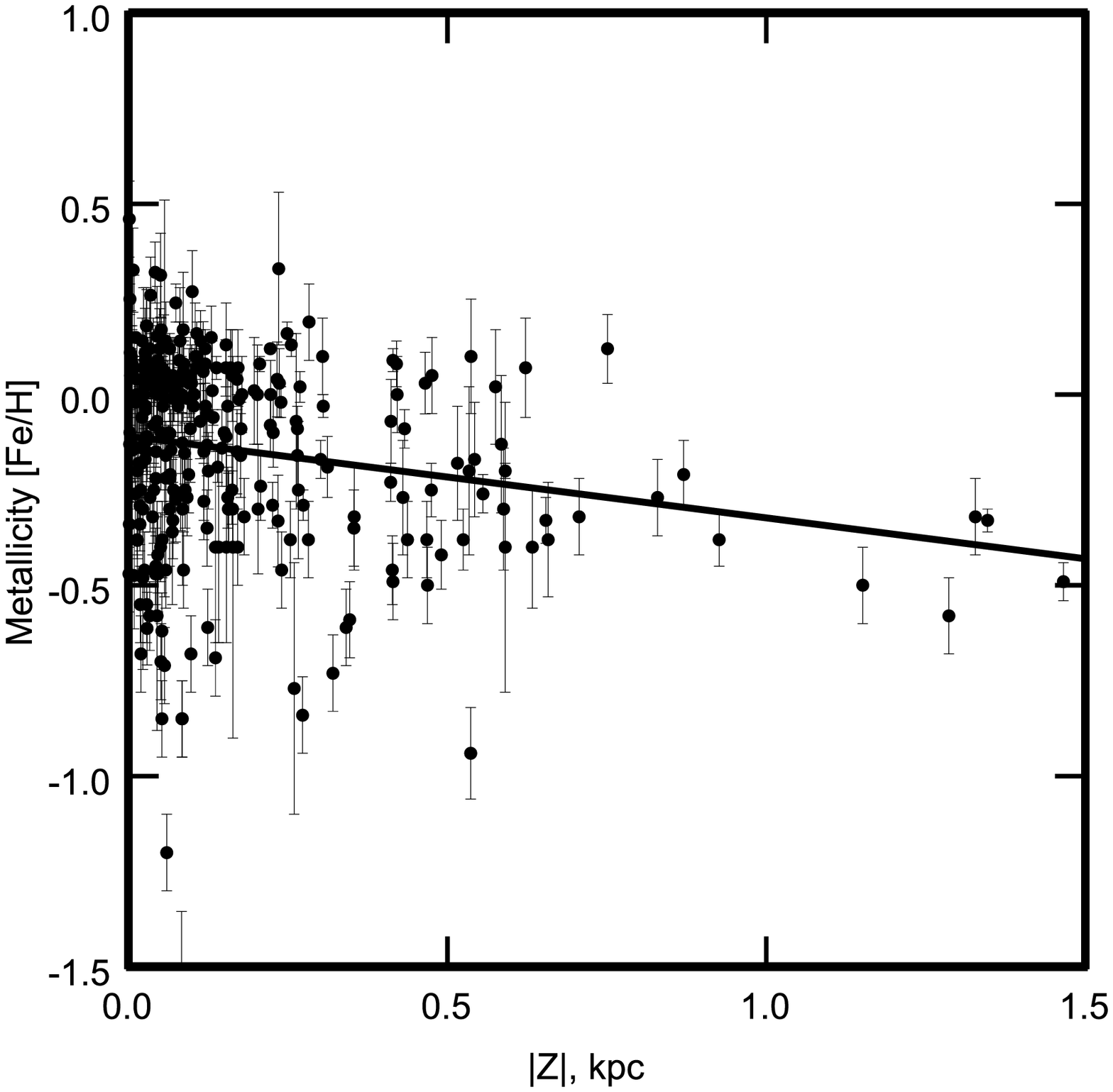}
  \end{minipage}%
  \caption{Dependencies of OCl metallicity on the galactocentric distances (left),
  decimal logarithm of age (middle) and the distance from the Galactic plane (right).
  Solid lines show regression models (2), (5) and (6), respectively. Vertical bars
  show the errors of metallicities in the catalogs used.}
  \label{Fig2}
  \end{figure}

Figure~\ref{Fig2} (left) shows the metallicity estimate dependence
on the galactocentric distances, R, for the OCls of our sample.
Galactocentric distances are calculated using the galactocentric
distance of the Sun $8.3$~kpc ~\citep{Gerasimenko2004}. To present
the available data, we tested the following regression models:

(1) the simplest model, with no gradient. That is, the metallicity
is equal to the average metallicity of the sample clusters
$<[Fe/H]> = -0.147 (\pm 0.270)$. In Figure~\ref{Fig2} (left), a
horizontal dashed line shows this model, which is clearly
inadequate for extreme galactocentric distance values.

(2) linear dependence of metallicity on galactocentric distances,
with the coefficient values  determined by the least squares
method:

$$
 [Fe/H] = 0.241 (\pm 0.065) - 0.041 (\pm 0.007)\cdot R.
$$

In Figure~\ref{Fig2} (left), the corresponding regression line is
shown as solid line. The coefficient at R is the linear gradient
of metallicity in the Galactic disk. The standard deviation of
residuals from regression model is equal to 0.25. The average
error in the catalog values for $[Fe/H]$ is 0.1, however, we use
heterogeneous material, making it difficult to examine the reasons
for the large dispersion of OCl metallicities. Testing the
statistical significance of the inclusion of a linear R term in
the model gives a dispersion ratio of $1.14$ for models (1) and
(2), which, according to the Fisher distribution, indicates that
the linear term is significant for our sample size and a $95 \%$
significance level.

(3) quadratic model for galactocentric distance:

$$
  [Fe/H] = 0.300 (\pm 0.242) - 0.053 (\pm 0.048)\cdot R + 0.001 (\pm 0.002)\cdot R^2.
$$

The ratio of the coefficient of the squared galactocentric
distance and its error hints at the statistical insignificance of
this regression model term. Indeed, the dispersion ratio is
$1.003$ for models (2) and (3), and, according to the Fisher
distribution value, the quadratic term is not significant at any
reasonable significance level value.

(4) linear model, adjusted for the dependence on time (OCl age):

$$
  [Fe/H] = -0.258 (\pm 0.167) - 0.049 (\pm 0.007)\cdot R + 0.069 (\pm 0.021)\cdot \log T.
$$

The dispersion ratio is equal to $1.03$ for models (2) and (4),
indicating that the time term in the model is statistically
insignificant.

(5) purely time dependence of the metallicity:

$$
  [Fe/H] = -0.264 (\pm 0.174) + 0.014 (\pm 0.021)\cdot \log T.
$$

Figure~\ref{Fig2} (middle) shows this dependence. The standard
deviation of residuals from regression model is equal to 0.27. The
dispersion ratio is $1.001$ of models (1) and (5), and the
dependence is statistically insignificant.

(6) - linear dependence on the absolute value of the distance from
the Galactic plane, $|Z|$:

$$
  [Fe/H] = -0.109 (\pm 0.019) - 0.214 (\pm 0.065)\cdot |Z|.
$$

This model was considered due to the fact that some of the oldest
OCls may belong to the thick disk of the Galaxy and have, on
average, less metallicity than clusters in the thin disk.
Figure~\ref{Fig2} (right) shows the dependence of metallicity on
the OCl distance from the Galactic plane. The standard deviation
of residuals from regression model is equal to 0.26. The
dispersion ratio is $1.04$ for models (1) and (6), that is, the
regression of $|Z|$ is statistically insignificant.

\section{Nonlinearity of dependence of metallicity on Galactocentric distance}
\label{sect:nonlinear}

\begin{figure}
   \centering
   \includegraphics[width=7cm, angle=0]{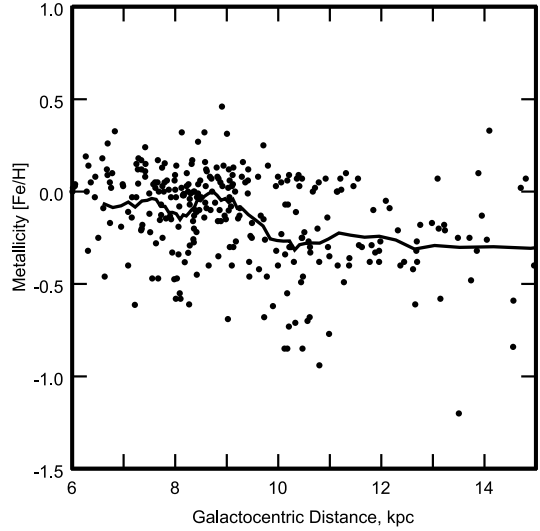}
   \caption{The smoothed dependence of OCl metallicities on the galactocentric distances.}
   \label{Fig3}
   \end{figure}

We established that the quadratic term in the regression model of
the dependence of metallicity on the galactocentric distance is
not statistically significant. In previous studies, instances of
nonlinearity in this dependence appear regularly (see the
references above). Figure~\ref{Fig3} shows the smoothed dependence
between the metallicity of all the sample clusters and the
galactocentric distance, with points representing individual
clusters. Smoothing was performed in two steps: (1)~smoothing with
a 9-point digital low-pass filter; (2)~and sliding interval
averaging for groups of 20 points. The stepwise nature of the
resulting dependence is clearly visible: a distinct jump in the
average OCl metallicity is observed in the region of $R = 9 - 10$~
kpc. Figure~3 in \cite{Twarog1997} also clearly demonstrates this
jump, as does figure~2a in \cite{Gozha2012C}. Interestingly, this
jump is practically unobservable in the corresponding dependence
for classical Cepheids. The reasons for this jump are difficult
explain, but it must be taken into account when building models of
the chemical evolution of the Galaxy.

We decided to examine the internal and external regions of the
Galaxy, with respect to the jump, separately. Of all the clusters
in the sample, we selected two groups of clusters according to
their distances from the center of the Galaxy, $R < 9$~kpc and $R
> 10$~kpc. For both groups, regression lines were obtained for the
linear dependencies between metallicities and galactocentric
distances, the results of which are presented in the third column
of Table~\ref{Tab1} (first and third lines). The last column of
Table~\ref{Tab1} shows the corresponding dispersion ratios. We
found that the slopes of the lines are statistically
insignificant. Therefore, the ratio of metallicity dispersion to
the deviation dispersion from the approximating straight line is
$1.018$ and $1.006$ for clusters in the inner region of the Galaxy
($R < 9$~kpc) and in the outer ($R > 10$~kpc) region,
respectively.

\begin{table}
\begin{center}
\caption[]{ Regression Lines for the Dependence of OCl Metallicity
on the Galactocentric Distance and Distance from the Galactic
Plane.}\label{Tab1}


 \begin{tabular}{lccc}
  \hline\noalign{\smallskip}
 & Sample size & $[Fe/H]$ & Dispersion Ratio  \\
  \hline\noalign{\smallskip}
 $R < 9$ kpc   & 153 & $-0.067 (\pm 0.177) + 0.0002 (\pm 0.023) \cdot R$  & 1.018  \\
               &     & $-0.068 (\pm 0.026) - 0.003 (\pm 0.215) \cdot |Z|$ & 1.007  \\
 $R > 10$ kpc  &  99 & $-0.144 (\pm 0.201) - 0.012 (\pm 0.017) \cdot R$   & 1.006  \\
               &     & $-0.255 (\pm 0.040) - 0.087 (\pm 0.087) \cdot |Z|$ & 1.001  \\
  \noalign{\smallskip}\hline
\end{tabular}
\end{center}
\end{table}

Thus, as previously noted by \cite{Twarog1997}, the dependence of
metallicity on the galactocentric distance is represented as two
horizontal line segments spaced along $[Fe/H]$ at intervals of
$0.22$, with average metallicities of $-0.065 (\pm 0.019)$ for $R
< 9$~kpc and $-0.284 (\pm 0.028)$ for $R > 10$~kpc.

\begin{figure}
   \centering
   \includegraphics[width=7cm, angle=0]{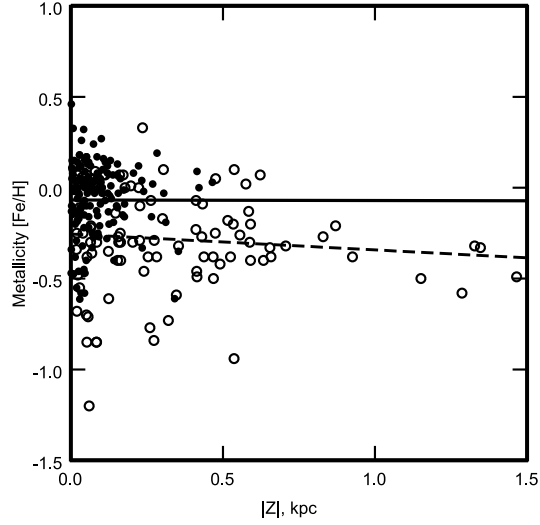}
   \caption{Dependence of OCl metallicities on distance
   from the Galactic plane, with regression lines for distances of $R < 9$~kpc
   (black circles, solid line) and $R > 10$~kpc (open circles, dashed
   regression line).}
   \label{Fig4}
   \end{figure}

We decided to reexamine the dependence of metallicity on the
distance of OCls from the Galactic plane separately for the
internal and external regions, with respect to the metallicity
jump. Table~\ref{Tab1} (second and fourth lines) and in
Figure~\ref{Fig4} show the results of attempts to approximate by
straight lines. The last column of Table~\ref{Tab1} indicates the
insignificance of the slopes of these dependencies, therefore, we
did not get any statistically significant metallicity dependencies
for $|Z|$. If the absence of this dependence is expected for $R >
10$~kpc since there are practically no metal-rich thin disk
clusters in this region, then its absence for $R < 9$~kpc is most
likely indicates insufficient data, as thick disk clusters growing
in proportion with $|Z|$ could lead to the occurrence of such a
dependence.

\section{Generations}
\label{sect:gener}

Above, we examined the average metallicity dependency on cluster
age and found that both the inclusion of an age-dependent term in
the general model (see model (5) above) and a separate analysis of
the age dependence yielded no definite results. We decided to try
a slightly different approach to resolve this issue. In
\cite{Popova2008}, we identified the individual generations of
OCls that belong to regions of different spiral arms. It can be
assumed that the metallicities of different OCls within the volume
of one space and one generation should be close. In this case, we
can average the OCl metallicities of one generation and one space
volume, which should lead to more reliable estimates in regard to
random errors. Our sample clusters were divided into groups
according to their spiral arms and generations: the
Carina-Sagittarius arm (Car-Sag) with R about $7.03$~kpc, three
generations; the Orion arm (Ori) with R about $8.80$~kpc, three
generations; the Perseus arm (Per) with R about $10.49$~kpc, two
generations. Arm positions were taken from \cite{Popova2005}. For
each generation of each arm, the average $\log T$ and average
metallicity $[Fe/H]$ were determined along with the corresponding
standard errors. The results are shown in Table~\ref{Tab2}.

\begin{table}
\begin{center}
\caption[]{ Dependence of the Average OCl Metallicity on the
Average Decimal Logarithm of Age.}\label{Tab2}


 \begin{tabular}{ccccc}
  \hline\noalign{\smallskip}
Spiral Arm & Generation & $<\log T>$ & $<[Fe/H]>$ & Sample size  \\
  \hline\noalign{\smallskip}
Car-Sag & 1 & 7.35 $\pm$ 0.16 & -0.116 $\pm$ 0.273 & 10  \\
Car-Sag & 2 & 8.08 $\pm$ 0.15 & -0.037 $\pm$ 0.193 & 17  \\
Car-Sag & 3 & 9.01 $\pm$ 0.12 & -0.070 $\pm$ 0.187 &  9  \\
Ori     & 1 & 7.01 $\pm$ 0.10 & -0.161 $\pm$ 0.306 &  9  \\
Ori     & 2 & 7.65 $\pm$ 0.14 & -0.040 $\pm$ 0.168 & 19  \\
Ori     & 3 & 8.67 $\pm$ 0.17 & -0.063 $\pm$ 0.217 & 37  \\
Per     & 1 & 8.58 $\pm$ 0.08 & -0.198 $\pm$ 0.197 &  8  \\
Per     & 2 & 9.14 $\pm$ 0.09 & -0.202 $\pm$ 0.172 & 11  \\
  \noalign{\smallskip}\hline
\end{tabular}
\end{center}
\end{table}

\begin{figure}
   \centering
   \includegraphics[width=7cm, angle=0]{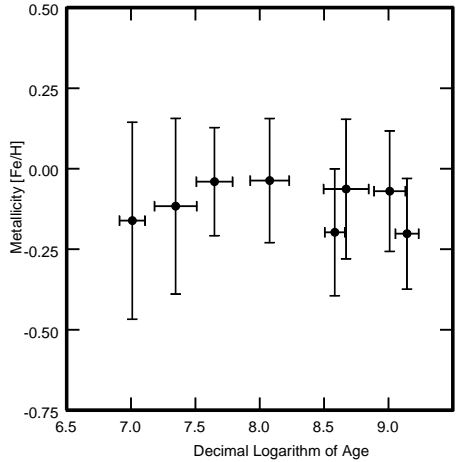}
   \caption{Dependence of the average OCl metallicity on the average
   logarithm of age. Each point represents a separate
   generation of an individual spiral arm.}
   \label{Fig5}
   \end{figure}

The values from Table~\ref{Tab2} are shown in Figure~\ref{Fig5}.
Each point represents a separate generation for each of the spiral
arms. This figure does not show any noticeable age trend in the
average OCl metallicity, which leads us to acknowledge the fact
that the present data do not allow us to discuss not only the rate
of heavy-element enrichment of the Galactic disk, but even the
observability of the enrichment itself.

\section{Summary and conclusions}
\label{sect:discussion}

The chemical composition gradient of the galactic disk was
examined based on a sample of OCls with available metallicity
estimates and improved distance estimates provided by Gaia DR2. As
reported by \cite{Twarog1997}, we did not see a monotonic
variation in the average OCl metallicity as the galactocentric
distance increased. This dependence clearly shows a jump in the
metallicity $[Fe/H]$ of $0.22$ at a Galactocentric distance of
about $9.5$~kpc, which appears to be linked  to the outer boundary
of the thinnest component of the galactic disk. This supports the
idea of \cite{Gozha2012C} that OCls in the solar vicinity are two
subsystems that differ in kinematics and chemical composition. The
absence of a significant metallicity gradient in the inner ($R <
9$~kpc) and external ($R > 10$~kpc) regions of the disk shows the
absence of a noticeable metal enrichment during the ordering of
the ages that correspond to those of the disk regions under
consideration. Observational data show that a noticeable metal
enrichment of the disk occurs only during the starburst epochs,
when the galactic disk components, such as a thick and thin disk,
and maybe other not so clearly visible components form.
Unsuccessful attempts to obtain a dependence of average
metallicity on age also point toward this conclusion. Errors, both
systematic and random, in the estimates of distances naturally
lead to changes in the gradient of the chemical composition of the
Galaxy. Obviously, random errors in distance estimates on average
cause an overstatement of distances, which means an
underestimation of the chemical composition gradients. An attempt
to introduce statistical corrections of the influence of random
errors for distances \citep{Loktin2019} did not lead to noticeable
changes in the results, so we did not use these corrections in
this paper.

Unfortunately, the data on OCl metallicities are still rather
unreliable. One would expect that a large number of uniform
metallicity estimates would become available after implementation
the RAVE project \citep{Kordopatis2013}. Unfortunately, most of
OCls are concentrated near a plane of the Galaxy for which the
RAVE DR5 catalog has no information. Data from the LAMOST survey
\citep{Deng2012, Zhao2012}, especially the recently created DR5,
show great promise.

\begin{acknowledgements}
This work was supported in part by the Ministry of Education and
Science (the basic part of the State assignment, RK no.
AAAA-A17-117030310283-7) and by the Act no. 211 of the Government
of the Russian Federation, agreement no.02.A03.21.0006.
\end{acknowledgements}

\bibliographystyle{raa}
\bibliography{LoktinPopova}

\label{lastpage}

\end{document}